\documentclass[twocolumn, prb, a4paper, 10pt, superscriptaddress, nopacs, floatfix, longbibliography]{revtex4-1}
\usepackage{siunitx}
\usepackage{graphicx}
\usepackage{xcolor}
\usepackage{physics}
\usepackage{nicefrac}
\usepackage{bm}
\usepackage{float}
\usepackage[utf8]{inputenc}
\usepackage[colorlinks, citecolor=blue, linkcolor=black]{hyperref}
\usepackage{ulem}
\usepackage{booktabs}
\usepackage{multirow}
\usepackage{xspace}

\newcommand{\CT}{CrTe\textsubscript{2}\xspace}
\newcommand{\VB}{V$_\text{B}^-$\xspace}

\begin{document}

\title{Magnetic imaging under high pressure with a spin-based quantum sensor integrated\\
in a van der Waals heterostructure}

\author{Z. Mu}
\affiliation{Laboratoire Charles Coulomb, Universit\'e de Montpellier and CNRS, 34095 Montpellier, France}
\author{J.~Frauni\'e}
\affiliation{Universit\'e de Toulouse, INSA-CNRS-UPS, LPCNO, 135 Avenue Rangueil, 31077 Toulouse, France}
\author{A. Durand}
\author{S. Cl\'ement}
\author{A. Finco}
\affiliation{Laboratoire Charles Coulomb, Universit\'e de Montpellier and CNRS, 34095 Montpellier, France}
\author{J. Rouquette}
\affiliation{Institut Charles Gerhardt, Universit\'e de Montpellier and CNRS,
34095 Montpellier Cedex, France}
\author{A.~Hadj-Azzem}
\author{N.~Rougemaille}
\author{J.~Coraux}
\affiliation{Univ. Grenoble Alpes, CNRS, Grenoble INP, Institut NEEL, 38000 Grenoble, France}
\author{J. Li}
\author{T. Poirier}
\author{J.~H.~Edgar}
\affiliation{Tim Taylor Department of Chemical Engineering, Kansas State University, Kansas 66506, USA}
\author{I. C.~Gerber}
\author{X.~Marie}
\affiliation{Universit\'e de Toulouse, INSA-CNRS-UPS, LPCNO, 135 Avenue Rangueil, 31077 Toulouse, France}
\affiliation{Institut Universitaire de France, 75231 Paris, France}
\author{B. Gil}
\affiliation{Laboratoire Charles Coulomb, Universit\'e de Montpellier and CNRS, 34095 Montpellier, France}
\author{G. Cassabois}
\affiliation{Laboratoire Charles Coulomb, Universit\'e de Montpellier and CNRS, 34095 Montpellier, France}
\affiliation{Institut Universitaire de France, 75231 Paris, France}
\author{C. Robert}
\affiliation{Universit\'e de Toulouse, INSA-CNRS-UPS, LPCNO, 135 Avenue Rangueil, 31077 Toulouse, France}
\author{V. Jacques}
\email{vincent.jacques@umontpellier.fr}
\affiliation{Laboratoire Charles Coulomb, Universit\'e de Montpellier and CNRS, 34095 Montpellier, France}

\begin{abstract}
Pressure is a powerful thermodynamic parameter for tuning the magnetic properties of van der Waals magnets owing to their weak interlayer bonding. However, local magnetometry measurements under high pressure still remain elusive for this important class of emerging materials. Here we introduce a method enabling {\it in situ} magnetic imaging of van der Waals magnets under high pressure with sub-micron spatial resolution. Our approach relies on a quantum sensing platform based on boron-vacancy (\VB) centers in hexagonal boron nitride (hBN), which can be placed in atomic contact of any type of two-dimensional (2D) material within a van der Waals heterostructure. We first show that the \VB center can be used as a magnetic field sensor up to pressures of a few GPa, a pressure range for which the properties of a wide variety of van der Waals magnets are efficiently altered. We then use \VB centers in a thin hBN layer to perform magnetic imaging of a van der Waals magnet under pressure. As a proof of concept, we study the pressure-dependent magnetization in micrometer-sized flakes of $1T$-\CT, whose evolution is explained by a shift of the Curie temperature. Besides providing a new path for studying pressure-induced phase transitions in van der Waals magnets, this work also opens up interesting perspectives for exploring the physics of 2D superconductors under pressure via local measurements of the Meissner effect.
\end{abstract} 
\date{\today}

\maketitle

Magnetic van der Waals materials and their associated heterostructures are currently attracting conside\-rable interest, both for the study of magnetic order in two-dimensional (2D) systems and for the design of atomically-thin spintronic devices~\cite{Gibertini2019,Genome2022,Naturereview2022}. 
Due to their lamellar structure with weak interlayer bonding, the properties of van der Waals magnets are very sensitive to pressure. For example, a transition from antiferromagnetic to ferromagnetic interlayer coupling is achieved for thin exfoliated flakes of CrI$_3$ at a pressure of few GPa, owing to a structural transition from monoclinic to rhombohedral layer stacking~\cite{Song2019,Li2019}. Pressure can also be used to tune the Curie temperature as well as the balance between exchange interactions and magneto-crystalline anisotropies in a broad variety of van der Waals magnets~\cite{APL2018,PhysRevB.99.180407,PhysRevMaterials.2.051004,FGT2020,FaugerasNanoLett23}. Such an efficient control of magnetic interactions with pressure opens up interesting prospects for the design of exotic magnetic phases in van der Waals heterostructures. However, these studies are hampered by a lack of experimental methods providing {\it in~situ} magnetic imaging of micrometer-sized samples under pressure, a notoriously difficult task in high pressure science and technology.\\
  \begin{figure}[b!]
  \centering
  \includegraphics[width = 8.5cm]{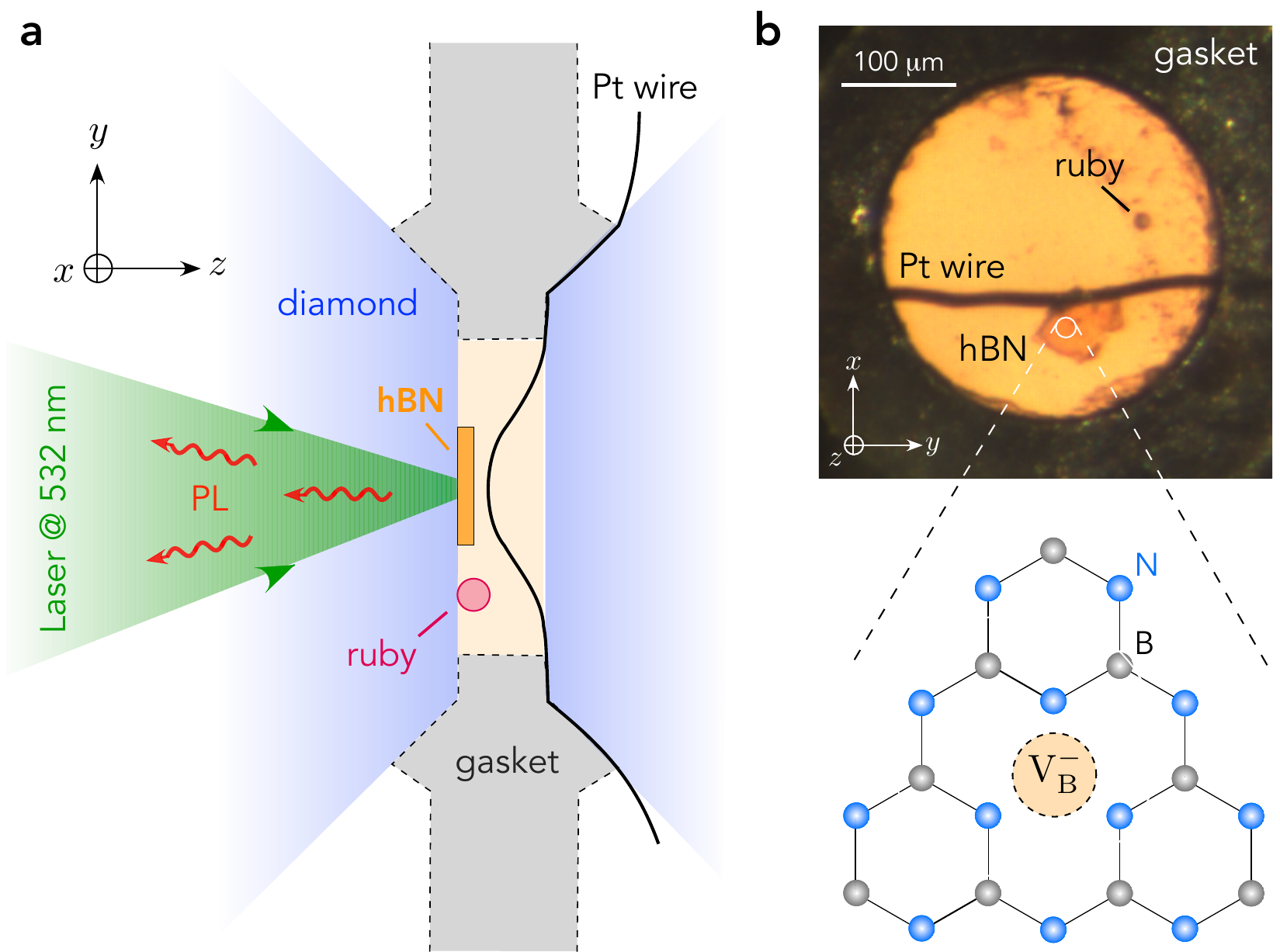}
  \caption{{\bf Principle of the experiment.} (a) Sketch of the experimental arrangement combining a membrane-driven DAC with a scanning confocal microscope. (b) Optical image of the sample chamber of the DAC showing a neutron-irradiated hBN crystal, a platinum (Pt) wire used for microwave excitation and a ruby crystal for local pressure measurements. Bottom panel: Atomic structure of the \VB center in hBN.}
  \label{fig1}
\end{figure}
A powerful approach to magnetic imaging under pressure with high sensitivity and sub-micron spatial resolution relies on the integration of a spin-based quantum sensor into a diamond anvil cell (DAC), which is the most widely used device to generate high hydrostatic pressure environments [Fig.~1(a)]. A prominent example is the use of nitrogen-vacancy (NV) centers implanted near the surface of a diamond anvil to probe magnetic signals generated inside the high-pressure chamber of a DAC via optically detected magnetic resonance (ODMR) methods~\cite{Doherty2014,Lesik2019,Yao2019,Yip2019,Shang_2019,Wang2024,Bhattacharyya2024}. Here we explore an alternative solution based on boron-vacancy (\VB) centers in hexagonal boron nitride (hBN), which is a promising platform for magnetic imaging in van der Waals heterostructures~\cite{Gottscholl2020,Du2021,PhysRevApplied.18.L061002,Tetienne2023}. Similar to the NV defect in diamond, the \VB center in hBN features a spin triplet ground state whose electron spin resonance frequencies can be interrogated by ODMR spectroscopy~\cite{Gottscholl2020}, a key resource for quantum sensing and imaging technologies.~Importantly, \VB centers hosted in thin hBN layers can be easily integrated into van der Waals hete\-rostructures, providing atomic-scale proximity between the spin-based quantum sensor and any type of 2D material.

\indent In this work, we demonstrate that \VB centers in hBN can be used for magnetic ima\-ging in van der Waals heterostructures under high pressure with sub-micron spatial resolution.~The novelty of our results is twofold.~First, we analyze the impact of hydrostatic pressure on the magneto-optical properties of \VB centers hosted in a bulk hBN crystal. We show that the ODMR response of the \VB center is preserved up to pressures of a few GPa despite a strong pressure-induced enhancement of non-radiative decay rates in optical cycles. We infer a global hydrostatic shift of the \VB center's electron spin resonance frequencies of~$\sim48$~MHz/GPa, a value in very good agreement with recent first-principles calculations~\cite{Udvarhelyi2023}. Second, we integrate \VB centers into a van der Waals heterostructure involving a 2D magnet and we demonstrate {\it in situ} magnetic imaging under pressure. As a proof of concept, we study the pressure-dependent magnetization in micrometer-sized flakes of $1T$-\CT, a van der Waals ferromagnet with in-plane magnetic anisotropy and a Curie temperature $T_{\rm c}\sim 320$~K at ambient pressure~\cite{Freitascrte22015,Sun2020,Coraux2020,PhysRevMaterials.5.034008}.
\vspace{0.2cm}

\indent We start by studying how the optical and spin properties of \VB centers in hBN evolve with pressure. To this end, we rely on a monoisotopic h$^{10}$BN crystal in which \VB centers are created through neutron irradiation~\cite{Liu2018,Li2021,haykal2021decoherence} (Methods). As sketched in Fig.~1(a), a piece of this bulk hBN crystal is loaded into the sample chamber of a membrane-driven DAC, which is defined by a hole drilled in a rhenium gasket that is compressed by two opposing diamond anvils. The sample chamber is filled with a pressure-transmitting medium, consisting of  NaCl or Daphne oil~\cite{Klotz_2009}, to provide a quasi-hydrostatic pressure environment and a ruby crystal is used as a local pressure gauge~\cite{Ruby} (see Supplementary Figure 1). The optical response of \VB centers under pressure is studied at room temperature with a scanning confocal microscope employing a green laser excitation, a long-working distance objective with a numerical aperture ${\rm NA}=0.42$ and a photon counting module (Methods). To perform ODMR spectroscopy, a microwave excitation is applied through a platinum wire crossing the DAC chamber [Fig.~1(b)].
 \begin{figure}[t!]
  \centering
  \includegraphics[width = 8.5cm]{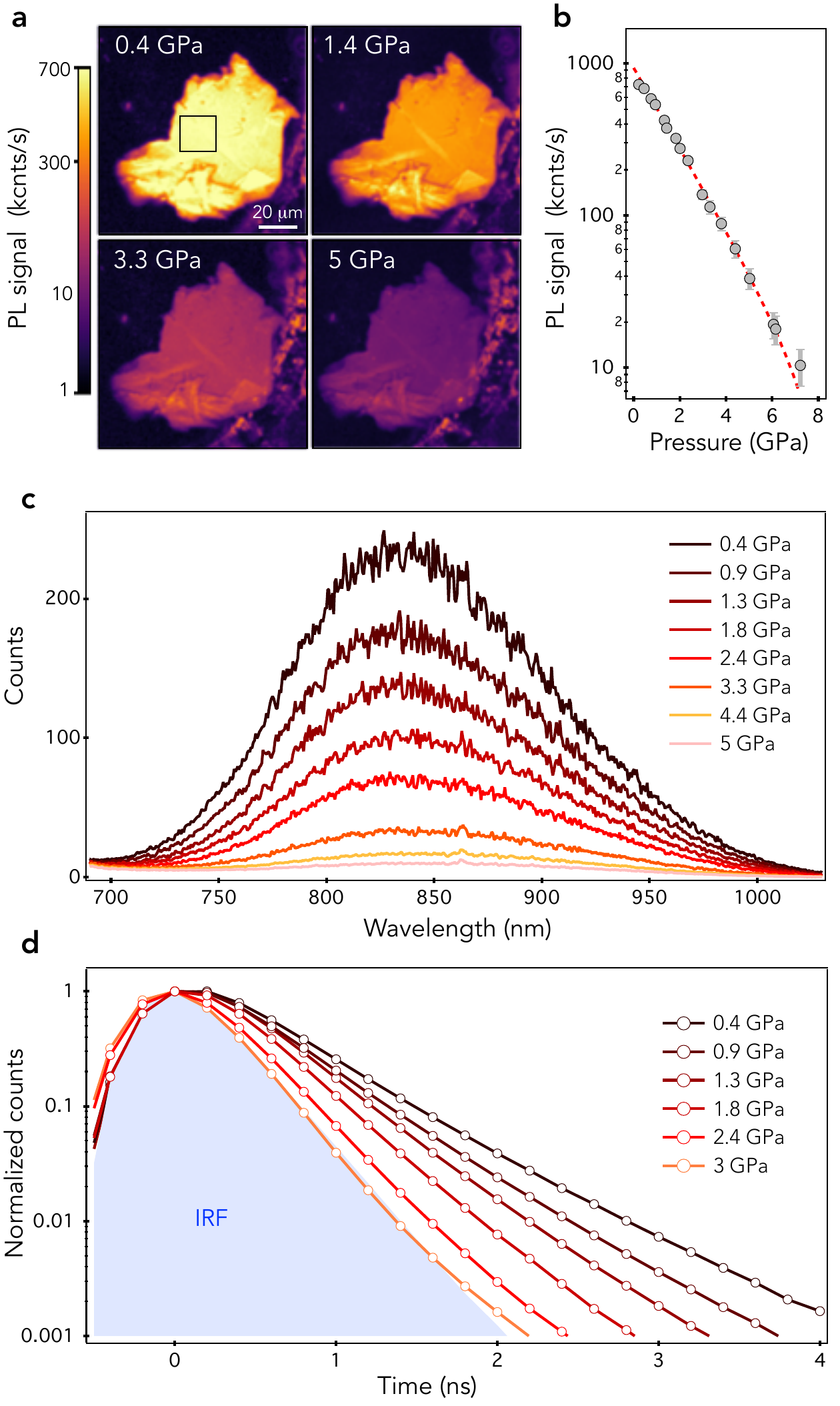}
  \caption{{\bf Optical properties of \VB centers in hBN under hydrostatic pressure.} (a) PL raster scans of the hBN crystal recorded at different pressures using a green laser power of $1$~mW. (b) PL signal averaged within the black square shown in (a) as a function of pressure. (c) Series of PL spectra and (d) time-resolved PL decays of \VB centers recorded at different pressures. The blue shaded area in (d) corresponds to the instrument response function (IRF) of the photon counting system measured with $60$-ps laser pulses.}
  \label{fig2}
\end{figure}

\indent Photoluminescence (PL) raster scans of the neutron-irradiated hBN crystal recorded at different pressures are shown in Fig.~2(a). The PL signal of \VB centers is drastically reduced with increasing pressure and becomes almost undetectable above $7$~GPa [Fig.~2(b)]. This behavior is confirmed by measurements of PL spectra, which feature the characteristic broadband emission of \VB centers in the near infrared with an intensity decreasing with pressure [Fig.~2(c)]. The process is reversible, the PL signal being recovered when pressure is released. To gain insight into the mechanism underlying this pressure-induced PL quenching, the lifetime of the \VB center's excited state was measured using a pulsed laser source. After optical excitation, the relaxation of \VB centers is dominated by non-radiative decay channels involving intersystem crossing (ISC) to a metastable state. These processes lead to an excited-state lifetime on the order of $\sim1$~ns under ambient conditions~\cite{baber2021excited,Clua2024,Whitefield2023}. Time-resolved PL decays recorded at different pressures show that the excited-state lifetime is significantly reduced with increasing pressure [Fig.~1(d)]. A quantitative analysis of this effect can hardly be performed since the PL decay of \VB centers cannot be distinguished from the instrument response function (IRF) of our photon counting module ($\sim 300$~ps) for pressure above 2 GPa. However, these measurements clearly indicate that pressure-induced PL quenching is linked to an enhancement of non-radiative decay rates in the optical cycles of \VB centers. 

We then investigate the impact of hydrostatic pressure on the spin triplet ground state ($S=1$) of the \VB center, whose Hamiltonian is expressed in frequency units as
\begin{equation}
\hat{\mathcal{H}}=D_0\hat{S}_z^2 + E(\hat{S}_x^2-\hat{S}_y^2) + \hat{\mathcal{H}}_{\rm s}\ ,
\end{equation}
where $\{\hat{S}_x,\hat{S}_y,\hat{S}_z\}$ are the dimensionless electron spin operators. The first term describes the spin-spin interaction that leads to an axial zero-field splitting $D_0\sim3.47$~GHz at ambient pressure between the $|m_s=0\rangle$ and $|m_s=\pm1\rangle$ spin sublevels~\cite{Gottscholl2020}, where $m_s$ denotes the electron spin projection along the $c$ axis~($z$) of the hBN crystal. The second term results from the coupling of the \VB center with a local electric field produced by surrounding charges~\cite{Gong2023NatCom,Udvarhelyi2023,DurandPRL2023}. As sketched in Fig.~3(a), this interaction mixes the $|m_s=\pm1\rangle$ spin sublevels, leading to
new eigenstates~$|\pm \rangle$ separated by an orthorhombic splitting $2E$, whose amplitude is linked to the volumic density of charges in the hBN crystal that increases with the density of \VB centers~\cite{Gong2023NatCom,Udvarhelyi2023}. Finally, the third term $\mathcal{H}_{\rm s}$ describes the spin-stress coupling. Considering a hydrostatic pressure environment, the stress tensor is diagonal with identical principal components $\sigma_{xx}=\sigma_{yy}=\sigma_{zz}=P$, where $P$ is the applied pressure. The electron spin density of the \VB center is mainly localized in the $(x,y)$ plane~\cite{ivady2020,Gracheva2023}. Since the stress component $\sigma_{zz}$ has a small impact on the in-plane hBN structure, its coupling to the \VB center can be safely neglected. In this framework, the Hamiltonian term describing the effect of hydrostatic pressure can be simply written as~\cite{Udvarhelyi2023}
\begin{equation}
\hat{\mathcal{H}_{\rm s}}=\alpha (\sigma_{xx}+\sigma_{yy})\hat{S}_z^2 = 2\alpha P\hat{S}_z^2\ ,
\end{equation}
where $\alpha$ denotes the stress coupling coefficient. Hydrostatic pressure thus leads to a global shift of the $|\pm \rangle$ spin sublevels, resulting in an increase of the zero-field splitting parameter $D=D_0+2\alpha P$ [Fig.~3(a)]. \\
  \begin{figure}[t!]
  \centering
  \includegraphics[width = 8.6cm]{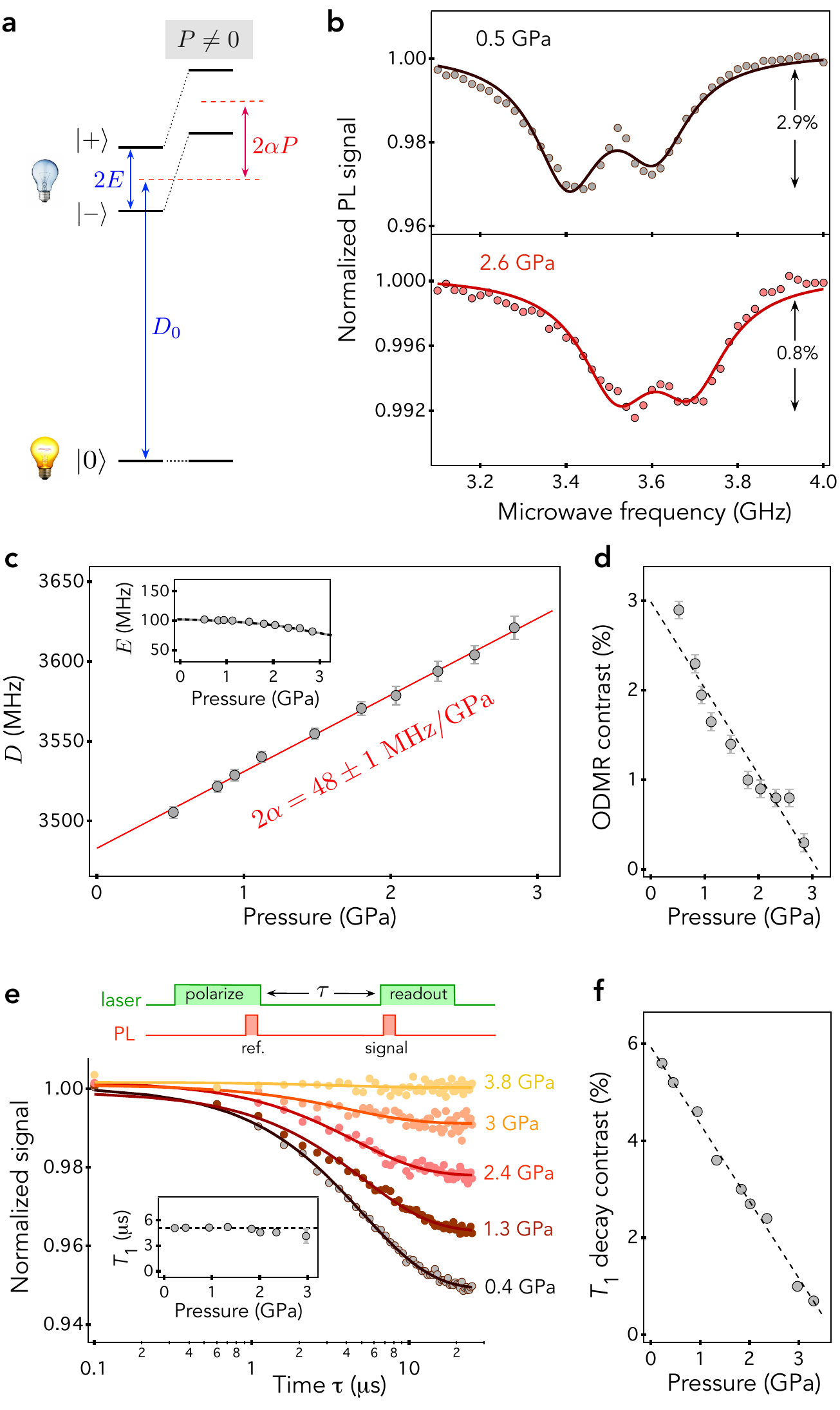}
  \caption{{\bf Spin properties under hydrostatic pressure.} (a) Energy level structure of the spin triplet ground state illustrating the impact of hydrostatic pressure $P$. (b) ODMR spectra recorded at $0.5$~GPa (top) and $2.6$~GPa (bottom). The solid lines are data fitting with Lorentzian functions from which the parameter $D$ and $E$ are extracted.
 (c) Zero-field splitting parameter $D$ as a function of $P$. The solid line is a fit with a linear function yielding $2\alpha=48\pm1$~MHz/GPa. Inset: Evolution of the $E$-splitting parameter with $P$. (d) ODMR contrast as a function of $P$. (e) Optically-detected spin relaxation curves recorded at different pressures using the experimental pulse sequence shown on top. The solid lines are data fitting with an exponential decay (see Methods). Inset: Evolution of the spin relaxation time $T_1$ with $P$.  (f) Contrast of the spin relaxation decay as a function of $P$.
   }
\end{figure}
\begin{figure*}[t]
  \centering
  \includegraphics[width = 18cm]{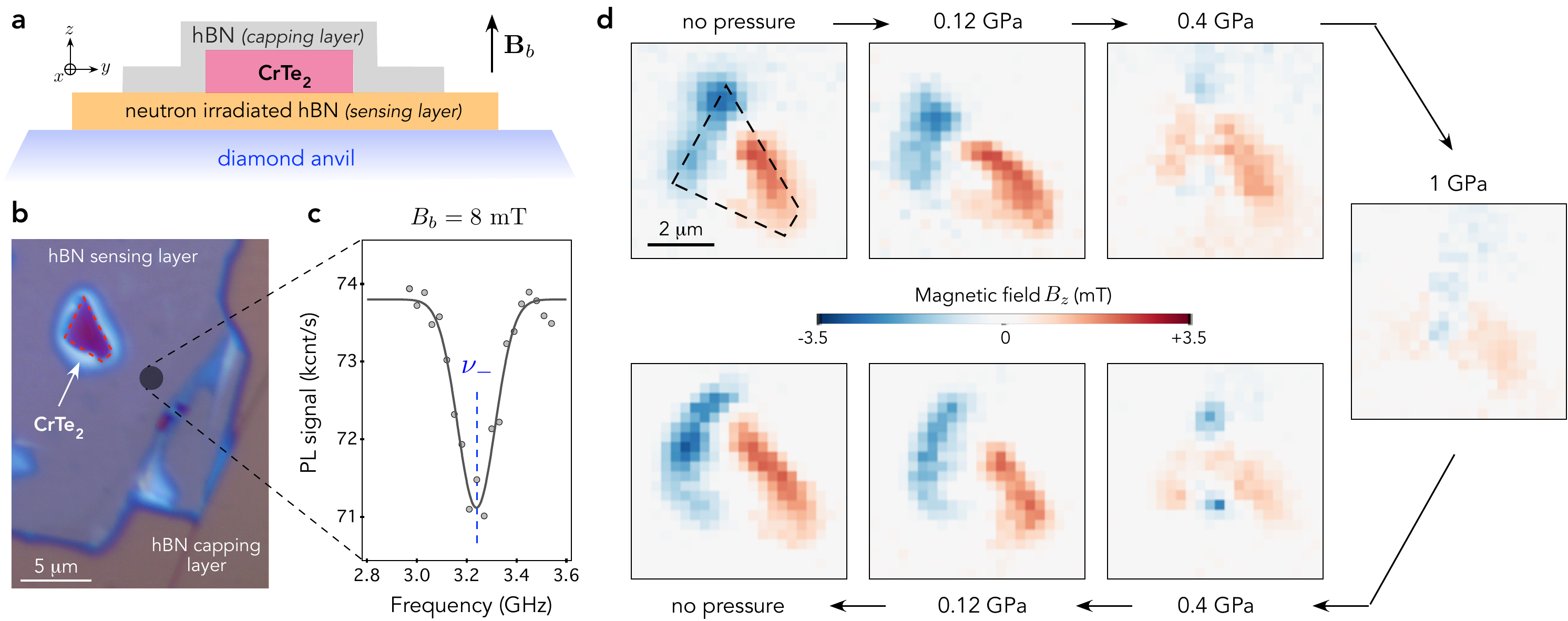}
  \caption{{\bf Magnetic imaging under pressure with \VB centers integrated in a van der Waals heterostructure.} (a) Sketch and (b) optical image of the heterostructure deposited in the DAC chamber. (c) Spectrum of the low-frequency magnetic resonance of the \VB center recorded far from the \CT flake at zero external pressure with a bias field $B_b=8$~mT. (d) Images of the magnetic field component $B_z$  produced by the \CT flake for increasing (top) and decreasing (bottom) pressure. For all images, the pixel size is $300\times 300$~nm$^{2}$ and the acquisition time per pixel is $60$~s. The pixel-to-pixel noise estimated in small regions outside the \CT flake is around $\sim 150 \ \mu$T, a value roughly in line with the estimation of the magnetic sensitivity using ODMR spectra. All experiments are performed at room temperature with a green laser power of $2$~mW.}
  \label{fig4}
\end{figure*}
\indent Typical ODMR spectra recorded at two different pressures are shown in Fig.~3(b). We detect the characteristic magnetic resonances of the \VB spin triplet ground state with frequencies given by $\nu_{\pm}=D\pm E$, which are shifted to higher frequency upon increasing pressure. The variation of the zero-field splitting parameter $D$ with pressure is plotted in Fig.~3(c). Data fitting with a linear function yields a stress coupling coefficient $2\alpha=48\pm1$~MHz/GPa. This value is in very good agreement with recent first-principles calculations predicting $2\alpha=49\pm2$~MHz/GPa for \VB centers embedded in a bulk hBN crystal~\cite{Udvarhelyi2023}. The  $E$-splitting remains almost unchanged though a slight reduction is observed at the highest applied pressures [see inset in Fig.~3(c)]. This effect, that cannot be explained by non-hydrostatic pressure components, is tentatively attributed to a decreased density of charges in the hBN crystal upon increasing pressure, which might result from pressure-induced charge state conversion of \VB centers. \\
\indent Besides the hydrostatic shift of the magnetic resonances, our measurements also reveal that the contrast of the ODMR signal decreases with pressure [Fig.~3(d)] and fully vanishes above $3$~GPa, a pressure at which the PL signal still remains easily detectable with a high signal-to-background ratio [Fig.~2(b)]. Such a reduction in ODMR contrast could simply result from a degradation of the platinum wire used for microwave excitation at high pressure. This hypothesis is excluded by microwave-free measurements of the longitudinal spin relaxation of \VB centers [Fig.~3(e) and Methods]. At low pressure, the optically-detected spin relaxation curve is highly contrasted ($\sim 5 \%$) and data fitting with an exponential decay yields a longitudinal spin decay time $T_1=5.1\pm0.1 \ \mu$s. When pressure increases, $T_1$ is not significantly modified while the contrast of the spin decay strongly decreases and vanishes completely above $3$~GPa [Fig.~3(f)]. These results confirm those obtained by ODMR spectroscopy. The spin readout contrast is linked to the combination of (i) an efficient polarization of the \VB center in the $\left|m_s=0 \right.\rangle$ ground state by optical pumping and (ii) a spin-dependent PL emission. These two effects result from the spin selectivity of non-radiative ISC transitions to and from a metastable state during optical cycles~\cite{baber2021excited,Clua2024,Whitefield2023}. While pressure-induced PL quenching discussed above is linked to an enhancement of non-radiative decay rates during optical cycles [Fig.~2(d)], the reduced ODMR contrast reveals a loss in spin selectivity of ISC transitions upon increasing pressure. 

Having established how the optical and spin properties of \VB centers in hBN evolve with hydrostatic pressure, we now turn to magnetic imaging under pressure with \VB centers integrated in a van der Waals heterostructure. As a proof of principle, we image the magnetic field produced by $1T$-\CT, a layered ferromagnet with in-plane magnetic anisotropy and a Curie temperature $T_{\rm c}\sim 320$~K under ambient pressure ~\cite{Freitascrte22015,Sun2020,Coraux2020,PhysRevMaterials.5.034008}. The van der Waals heterostructure deposited inside the high-pressure chamber of the DAC is described by Figs.~4(a,b). A thin hBN flake exfoliated from a neutron-irradiated crystal is first transferred on one of the diamond anvils and used as a magnetic detection layer. A micrometer-sized flake of \CT is then deposited on top of the sensing layer and capped with a hBN flake without \VB centers for protection. All layers of the heterostructure are few tens of nanometers thick. \\
\indent In what follows, a bias magnetic field $B_b=8$~mT is applied along the $z$ axis to split the electron spin resonances of the \VB center via the Zeeman effect
and we track the low-frequency component $\nu_-$ by ODMR spectroscopy. Neglecting the orthorhombic splitting parameter $E$, this frequency is given by $\nu_-(P)\approx D(P)-\gamma_e(B_b+B_z)$, where $\gamma_e=28$~GHz/T is the electron spin gyromagnetic ratio and $B_z$ denotes the magnetic field produced by the \CT flake along the $z$ axis. An ODMR spectrum recorded far from the ferromagnetic layer ($B_z=0$) without applying external pressure is shown in Fig.~4(c). From such a spectrum, the shot-noise limited magnetic field sensitivity $\eta$ of the hBN-based sensing layer can be estimated by 
$\eta\approx 0.7 \times \frac{1}{\gamma_e}\times\frac{\Delta\nu}{\mathcal C\sqrt{\mathcal R}}$, where $\mathcal{R}$ is the rate of detected photons, $\mathcal{C}$ the ODMR contrast and $\Delta\nu$ the linewidth~\cite{PhysRevB.84.195204}. We obtain $\eta\sim 400 \ \mu$T/$\sqrt{\rm Hz}$ for a diffraction-limited laser excitation spot below $1 \ \mu$m$^2$. The same experiment performed at $P=1$~GPa leads to a degraded magnetic sensitivity $\eta\sim 800 \ \mu$T/$\sqrt{\rm Hz}$ due to the reduction in both spin readout contrast and PL signal (see Supplementary Figure 2). Although better sensitivities can be obtained with NV centers implanted in a diamond anvil~\cite{Bhattacharyya2024}, the key advantage of the hBN-based sensing unit is its ability to be integrated in a van der Waals heterostructure, offering atomic-scale proximity to the sample being probed.\\ 
\indent Magnetic field imaging is carried out by recording an ODMR spectrum at each point of a scan across the van der Waals heterostructure. By fitting the spectra, we infer the Zeeman shift of the spin resonance frequency, from which a map of the magnetic field component $B_z$ is obtained. Images of the stray field distribution generated by the \CT flake at different pressures are shown in Fig.~4(d). At low pressure, a magnetic field with reversed sign is mostly produced at opposite edges of the flake, as expected for a ferromagnetic material with uniform in-plane magnetization~\cite{PhysRevMaterials.5.034008}. As pressure increases, the amplitude of the stray magnetic field reduces and almost vanishes at $1$~GPa. The magnetic images also indicate a sudden variation of the stray field distribution around $0.4$~GPa, which suggests the formation of magnetic domains within the flake that results from pressure-induced variations of magnetic interactions. This process is reversible, the initial magnetic configuration of the \CT flake being restored when the applied pressure is decreased [Fig.~4(d)]. Such measurements, which could not be carried out with conventional magnetometry techniques based on superconducting quantum interference devices (SQUID) or inductively coupled coils, illustrate the potential of \VB centers in hBN for {\it in situ} magnetic imaging of van der Waals magnets under pressure. 

 \begin{figure}[t!]
  \centering
  \includegraphics[width = 8.6cm]{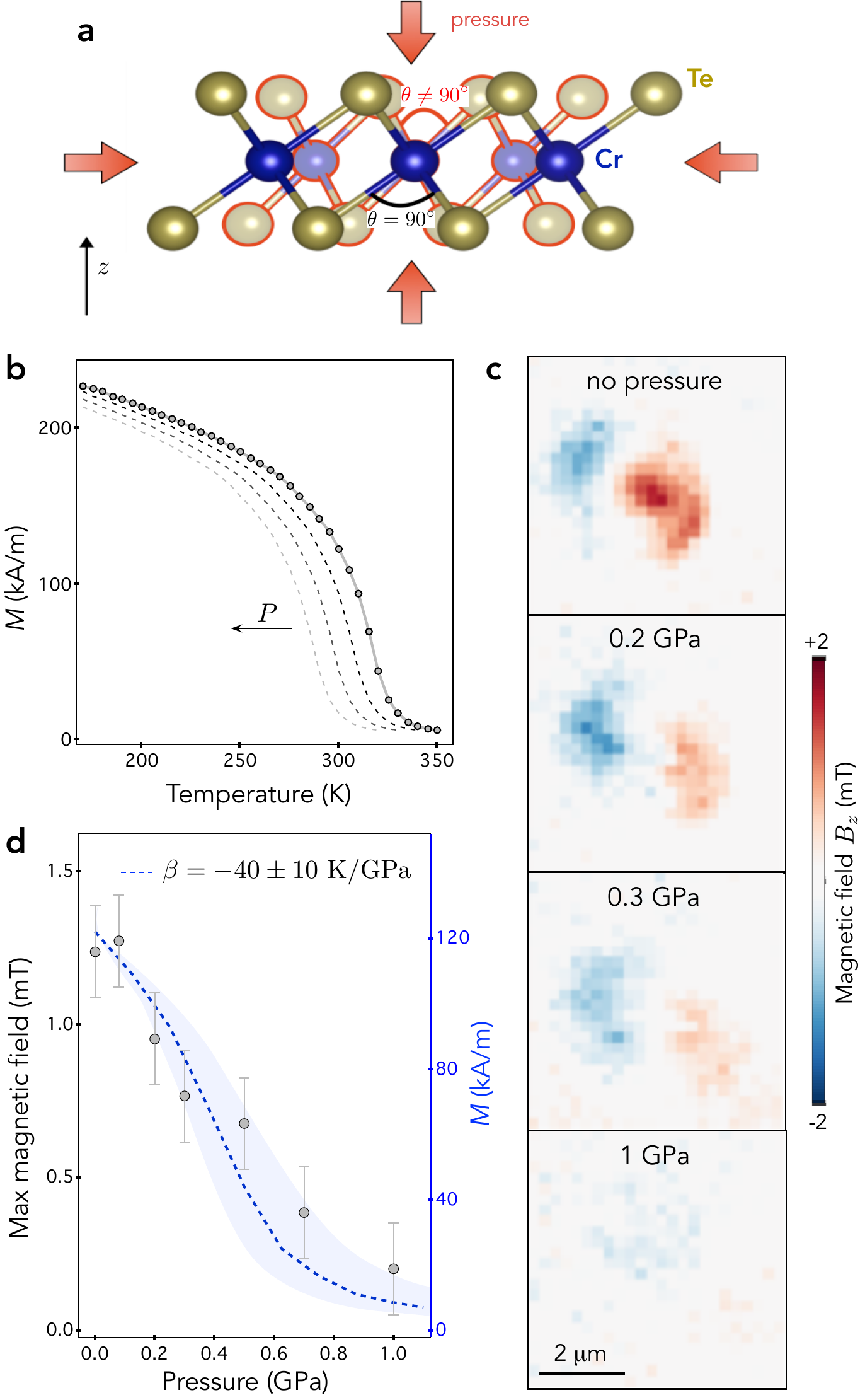}
  \caption{{\bf Pressure-induced shift of $\mathbf{T_c}$.} (a) Crystal structure of \CT illustrating the variation of the Cr-Te-Cr bond angle with applied pressure. (b) Temperature dependence of the magnetization $M$ of a bulk \CT crystal recorded by vibrating sample magnetometry.  The dotted lines correspond to artificial shifts in the data to account for a pressure-induced shift in Curie temperature. (c) Pressure-dependent magnetic field map of a second \CT flake recorded at room temperature. The pixel size is $250\times 250$~nm$^{2}$ and the acquisition time per pixel is $60$~s. (d) Maximum magnetic field measured at the edges of the \CT flake as a function of pressure. The blue shaded area correspond to the variation of $M$ with pressure at room temperature, which is obtained by using the data shown in (b) and a pressure-induced shift of the Curie temperature $\beta=-40\pm 10$~K/GPa.}
\end{figure}
\indent The pressure-dependent magnetization of \CT is explained by a reduction of the magnetic exchange interaction upon increasing pressure, which leads to a decreased Curie temperature. In \CT, the exchange interaction between nearest Cr neighbors is dominated by the superexchange coupling mediated by a Te ion. The strength of this interaction is very sensitive to the angle~$\theta$ of the Cr-Te-Cr bond. If $\theta=90^{\circ}$, the superexchange interaction is ferromagnetic with maximal coupling strength~\cite{APL2018,PhysRevB.99.180407}. First principles calculations of the cell parameters of \CT indicate that $\theta$ is very close to $90^{\circ}$ under ambient conditions (Methods). Upon increasing pressure, the angle $\theta$ decreases leading to a reduced strength of the superexchange coupling (see Supplementary Figure 3), which in turn results in a decreased Curie temperature. A similar effect was reported for the van der Waals magnet Cr$_2$Ge$_2$Te$_6$~\cite{APL2018} and for other Cr$_{1-\delta}$Te compounds~\cite{Ohta1993,Ishizuka2001}. To obtain an estimate of the pressure-induced shift of the Curie temperature, we rely on the empirical analysis described in Fig.~5(b). Using a measurement of the magnetization~$M$ of a bulk \CT crystal as a function of temperature, we first artificially infer its variations at room temperature by considering a pressure-induced shift of $T_c$ with a slope $\beta=dT_c/dP<0$ [Fig.~5(b)]. We then compare the resulting pressure-dependent magnetization curves $M_{\beta}(P)$ with the maximum magnetic field $B_z$ produced at the edges of a \CT flake, which is proportional to the magnetization. This analysis is done for a series of magnetic images recorded at different pressure on a second heterostructure [Fig.~5(c)]. For this \CT flake, we did not observe the formation of magnetic domains upon increasing pressure, making the comparison of the field at the edges more reliable. The results are shown in Fig.~5(d). Our simple analysis reproduces fairly well the experimental results for $\beta=-40\pm 10$~K/GPa.

\indent To conclude, we have shown that \VB centers in hBN constitute a promising quantum sensing platform for local magnetic imaging in van der Waals heterostructures under hydrostatic pressures up to few GPa, a pressure range for which the properties of a wide variety of 2D magnets are efficiently altered~\cite{Song2019,Li2019,APL2018,PhysRevB.99.180407,PhysRevMaterials.2.051004,FGT2020,FaugerasNanoLett23}. Besides providing a new path for studying pressure-induced control of  magnetic phases in 2D magnets, the hBN-based magnetic imaging unit also opens up interesting perspectives for exploring the physics of 2D superconductors~\cite{Review2Dsupra} under pressure via local measurements of the Meissner effect. Of particular interest could be the study of superconductivity in twisted bilayer graphene, whose critical temperature is very sensitive to pressures in the GPa range~\cite{TBLScience2019}. \\

\noindent {\it Note} - During the completion of this work, we became aware of a complementary study exploring the properties of \VB centers in hBN under pressure (Chong Zu, private communication), which will appear in the same arXiv posting~\cite{he2025probing}. 

\section*{Methods}
\noindent{\bf Diamond anvil cell (DAC).} All experiments were carried out with a membrane-driven DAC system equipped with two diamond anvils made of type-IIa single crystal diamonds with [100] crystalline orientation. The diamond anvils are cut into a 16-sided standard design with a culet diameter of $500 \ \mu$m (Almax-easyLab). The sample chamber is defined by a hole ($\sim250 \ \mu$m diameter) drilled into a rhenium gasket, which is squeezed between the two opposing diamond anvils. The high-pressure chamber of the DAC was filled with a pressure-transmitting medium to provide a quasi-hydrostatic pressure environment. For the study of the optical properties of \VB centers under pressure, we used Daphne oil 7373, while for the study of its spin properties and for magnetic imaging under pressure, NaCl was employed as pressure transmission medium. The fluorescence spectrum of a ruby microcrystal loaded into the DAC chamber was used as a local pressure gauge~\cite{Ruby}. In the pressure range considered in this work, we did not observe any broadening of the ruby emission lines (see Supplementary Figure 1), as expected for a quasi-hydrostatic pressure environment. 

To perform ODMR spectroscopy, a microwave excitation was delivered through a $10$-$\mu$m wide platinum wire crossing the DAC chamber. For these measurements, the rhenium gasket was coated with a mixture of epoxy and boron nitride powder to make it insulating in order to minimize microwave losses. \\

\noindent{\bf Sample preparation.} The hBN crystals studied in this work were synthesized by metal flux growth methods using a boron powder isotopically enriched with $^{10}$B~($99.2\%$) as precursor~\cite{Liu2018}. These crystals have a typical lateral size in the millimeter range and a thickness of a few tens of micrometers. Neutron irradiation was performed at the Ohio State University Research Reactor. The interest of isotopic purification with $^{10}$B lies in its very large neutron capture cross section, which ensures an efficient creation of \VB centers throughout the entire volume of the hBN crystal via neutron transmutation doping~\cite{Li2021}. 

The hBN crystal used to study the optical and spin properties of \VB centers under pressure was irradiated with a dose of~$\sim2.6 \times 10^{17}$~neutrons/cm$^{2}$. This crystal was cut into pieces of a few tens of micrometers, which were loaded inside the high-pressure DAC chamber. Magnetic imaging under pressure was performed with thin flakes exfoliated from a second hBN crystal, which was isotopically purified with $^{15}$N to obtain a slight reduction of the ODMR linewidth~\cite{EdgarAdvMat,Clua2023,Gong2024}. This crystal was irradiated with a dose of  $\sim1.4 \times 10^{17}$~neutrons/cm$^{2}$.

The van der Waals heterostructure described in the main text was assembled in a glove box, using \CT flakes  exfoliated from a bulk $1T$-\CT crystal synthesized following the procedure described in Ref.~[\onlinecite{Freitascrte22015}]. \\

\noindent{\bf Scanning confocal microscope.} The DAC system was incorporated into a home-built scanning confocal microscope operating under ambient conditions. A continuous laser excitation at $532$~nm was focused onto the sample placed in the DAC chamber through a long-working distance microscope objective with a numerical aperture NA $=0.42$ (Mitutoyo). The laser beam was scanned across the sample with a fast steering mirror (Newport, FSM-300) combined with a pair of telecentric lenses. The DAC system was placed on a translation stage equipped with a differential micrometer screw (Newport,  DM-13L), which was used to adjust the focus of the laser beam onto the sample with a sensitivity of about $100$~nm. The PL signal was collected by the same objective, focused in a single mode fiber and finally directed either to a spectrometer or to a silicon avalanche photodiode (APD) operating in the single photon counting regime (Excelitas, SPCM-AQRH). A $560$~nm long pass filter was used for measurements of PL spectra and an additional bandpass filter (Semrock, 935/170) was inserted for measurements of the PL signal with the APD. To infer the excited-state lifetime of the \VB center, we used a pulsed laser source at $531$~nm with a pulse duration of $60$~ps (PicoQuant, PDL-800D). Time-resolved PL measurements were obtained by recording photon detection events on a time tagger with a bin width of $200$~ps. 

For ODMR measurements, a microwave synthesizer was followed by a $15$~W power amplifier (Mini-circuits, ZHL-15W-422-S+) and connected to the platinum wire crossing the DAC chamber, which was terminated by a 50~$\Omega$ impedance. ODMR spectra were recorded by measuring the PL signal while sweeping the frequency of the microwave excitation.\\

\noindent{\bf $T_1$ measurements.} The longitudinal spin relaxation time ($T_1$) of the \VB center was measured with the experimental sequence sketched in the top panel of Fig.~3(e). A $5$-$\mu$s-long laser pulse produced with an acousto-optic modulator was first used to polarize the \VB centers in the ground state $|m_s=0\rangle$ by optical pumping. After relaxation in the dark during a variable time $\tau$, the remaining population in $|m_s=0\rangle$ was probed by integrating the spin-dependent PL signal produced at the beginning ($500$~ns window) of a second $5$-$\mu$s-long laser pulse. This signal was normalized with a reference PL value obtained at the end of the first laser pulse. The longitudinal spin relaxation time $T_1$ was then inferred by fitting the decay of the normalized PL signal with an exponential function of the form $1-C[1-\exp(-\tau/T_1)]$, where $C$ is the contrast of the optically-detected  spin decay. \\

\noindent{\bf First-principles calculations.} The atomic structure and the quasi-particle band structure of \CT have been obtained from DFT calculations using the VASP 
package \cite{Kresse:1993a,kresse:1996a}. The Perdew-Burke-Ernzerhof (PBE)~\cite{PBE1} functional was used as an approximation of the exchange-correlation electronic term. Additionally DFT-D correction is applied to take into account van der Waals interactions~\cite{Grimme:2010ij}. 
The plane-augmented wave scheme~\cite{blochl:prb:94,kresse:1999a} is used to treat core electrons, with twelve electrons for Cr atoms and six electrons for Te atoms in the valence states. To relax the structures upon hydrostatic pressure, the cell parameters are allowed to vary and all atomic positions are relaxed with a force convergence criterion below  $0.005$ eV/\AA. A grid of 21$\times$21$\times$9 $k$-points has been used, in conjunction with a Gaussian smearing with a width of 0.05 eV to determine partial occupancies. The cutoff energy of the plane-wave basis set was fixed to 500 eV. To determine magnetic states, the Hubbard U correction (PBE+U)~\cite{PhysRevB.57.1505} and spin-orbit coupling are included. The value of the U parameter was set as 5.8 eV as proposed in Ref.~[\onlinecite{PhysRevMaterials.6.084004}]. \\ 
\indent The calculated pressure-dependent lattice constants $a$ and $c$ of \CT are given in Supplementary Figure 3 for hydrostatic pressures up to 2 GPa. The $a$ and $c$ lattice parameters are reduced by around 1\% at 2 GPa. As a consequence the Cr-Cr distance is also decreasing, and the Cr-Te-Cr angle gets smaller than $90^{\circ}$ with increasing pressure. Based on a Heisenberg model, we can estimate the nearest neighbor exchange coupling with a Hamiltonian that can be expressed as $$\displaystyle \mathcal{\hat{H}}=-\sum_{ij} \hat{S}_{x}^{(i)} J \hat{S}_{x}^{(j)} +\sum_{i}A^{(i)} \left(\hat{S}^{(i)}_z\right)^2 \ ,$$ where $J$ is the effective exchange coupling constant, $\mathbf{\hat{S}}=(\hat{S}_x,\hat{S}_y,\hat{S}_z)$ is the spin vector, and $A^{(i)}$ is the magnetic anisotropy of a single ion. To calculate $J$ one can use energy difference between the ferromagnetic (FM) and antiferromagnetic (AFM) configurations simulated in $2\times2\times1$ supercell, since $\displaystyle E_{\rm FM}=E_o-12JS_x^2$ and $\displaystyle E_{\rm AFM}=E_o+4JS_x^2$, where $E_o$ is the total energy of systems without magnetic coupling.  At 0 GPa ,the magnetic moment of each Cr atom is 3.586 $\mu_B$, and the effective $J$ of monolayer is $2.24$~meV. When pressure is increased those values decrease (see Supplementary Figure 3).\\

\noindent {\it Acknowledgements} - The authors thank Manuel Nu\~{n}ez-Regueiro for valuable discussions. This work was supported by the French Agence Nationale de la Recherche through the program ESR/EquipEx+ 2DMAG (grant number ANR-21-ESRE-0025), the program PEPR SPIN (ANR-22-EXSP-0008), the projects Qfoil (ANR-23-QUAC-0003) and NEXT (ANR-23-CE09-0034), and the Institute for Quantum Technologies in Occitanie. hBN crystal growth was supported by the Materials Engineering and Processing program of the National Science Foundation, award number CMMI 1538127 and by the Office of Naval Research award N00014-22-1-2582. Neutron irradiation of the hBN crystal was supported by the U.S. Department of Energy, Office of Nuclear Energy under DOE Idaho Operations Office Contract DE-AC07-051D14517 as part of a Nuclear Science User Facilities experiment. I.C.G. acknowledges the computer resources through the "Calcul en Midi-Pyr\'en\'ees" initiative CALMIP (projects p0812) and CINES, IDRIS, and TGCC under the allocation 2024-A0160906649 made by GENCI, and the French Agence Nationale de la Recherche under grant agreement ANR-23-QUAC-004 (EXODUS) for financial support.


%

\end{document}